\documentclass[a4paper, amsfonts, amssymb, amsmath, reprint, showkeys, footinbib, twoside,superscriptaddress,floatfix,longbibliography]{revtex4-1}
\usepackage{tabularx}
\usepackage{booktabs}
\usepackage{multirow} 
\usepackage{graphicx}%
\usepackage{dcolumn}%
\usepackage{bm}%
\usepackage{float}
\usepackage{xcolor}
\usepackage{mhchem}
\usepackage{gensymb}
\usepackage[slantedGreek]{newtxmath}
\usepackage{comment}
\usepackage[pdftex, bookmarks, pdffitwindow=false, pdfstartview=FitH, pdfdisplaydoctitle, colorlinks, plainpages=false, pdftitle={},pdfauthor={}, pdfpagelabels, hypertexnames, citecolor={blue!50!black},linkcolor={blue!50!black}, urlcolor={blue!50!black}, pdflang={en}, hyperfootnotes=false, breaklinks]{hyperref}
\usepackage{siunitx}

\newcommand{\figLabelCapt}[1]{(\MakeLowercase{{#1}})}
\newcommand{\refSub}[2]{\hyperref[#2]{\ref{#2}\figLabelCapt{#1}}}
\newcommand{\figref}[1]{Fig.~\ref{#1}}
\newcommand{\figrefsub}[2]{Fig.~\refSub{#2}{#1}}
\newcommand{\figsrefsub}[2]{Figs.~\refSub{#2}{#1}}

\makeatletter
\renewcommand{\@seccntformat}[1]{}
\makeatother

\makeatletter
\def\@bibdataout@aps{
 \immediate\write\@bibdataout{
 @CONTROL{
   apsrev41Control, author="48",editor="1",pages="0",title="0",year="1"
 }}
 \if@filesw
  \immediate\write\@auxout{\string\citation{apsrev41Control}}
 \fi
}
\makeatother

\begin{document}
%\preprint{}

\title{Distilling latent electrostatics from foundation machine learning interatomic potentials} 

\author{Xiaoyu Wang}
\affiliation{Department of Chemistry, UC Berkeley, California 94720, United States}

\author{Bingqing Cheng}
\email{bingqingcheng@berkeley.edu}
\affiliation{Department of Chemistry, UC Berkeley, California 94720, United States}
\affiliation{Bakar Institute of Digital Materials for the Planet, UC Berkeley, California 94720, United States}
\affiliation{Chemical Sciences Division, Lawrence Berkeley National Laboratory, Berkeley, California, 94720, United States}

\date{\today}

\begin{abstract}
Foundation machine learning interatomic potentials (MLIPs) have enabled atomistic simulations across broad regions of chemical and materials space, but many remain computationally expensive and lack explicit electrostatics, limiting their use for systems governed by long-range interactions and electrical response. Previously, we introduced Latent Ewald Summation (LES), which learns latent atomic charges and long-range electrostatics from density functional theory (DFT) energy and force labels alone. Here, we use LES to extract electrostatics that are latent in foundation models: energies and forces predicted by a teacher model are used to train a lightweight LES-augmented student MLIP, with optional fine-tuning on additional DFT data. The resulting models reduce computational cost while providing access to Born effective charge tensors, and infrared spectra.
We benchmark student models distilled from a broad set of foundation MLIPs, including UMA, MACE, Orb, eSEN, GemNet-OC, PET, and EquiformerV2-based models, against experimental infrared spectra for liquid water, concentrated hydrochloric acid, and the anatase \ce{TiO2}(101)-water interface. Across these systems, electrostatic response can be extracted from most foundation MLIPs. The benchmark further shows that the underlying DFT level and dataset used to train the teacher model play a larger role than architecture in determining electrostatic and spectroscopic accuracy. For the \ce{TiO2}-water interface, fine-tuning with a modest amount of higher-level DFT data improves structural and infrared predictions. LES-based distillation therefore provides a practical route for converting foundation MLIPs into efficient, electrically responsive models, while also testing the physical fidelity encoded in foundation models.

\end{abstract}

\maketitle

\section{Introduction}
Machine learning interatomic potentials (MLIPs) are revolutionizing computational materials science and chemistry, by enabling fast and accurate atomistic simulations after training on quantum mechanical calculations such as density functional theory (DFT).
Foundation MLIPs are trained on large datasets across broad configurational and compositional spaces,
and are considered to be broadly applicable across the chemical space~\cite{Batatia2025foundationa, wood2025umafamilyuniversalmodels, sahoo2025open}.

However, current foundation MLIPs typically rely on increasingly expressive architectures with large parameter counts~\cite{wood2025umafamilyuniversalmodels,allam2025aqcat25, sahoo2025open, batatia2025crosslearning},
leading to higher computational costs that hinder large-scale downstream applications on specific chemical systems.
To overcome this limitation, recent studies have developed several strategies for knowledge distillation~\cite{hinton2015distilling} from expensive teacher MLIPs to cheaper student models. 
Beyond directly training on synthetic energy and force labels generated by teacher models~\cite{gardner2024synthetic}, recent work has explored the use of additional information from the teacher, such as latent atomic-energy decompositions for auxiliary supervision~\cite{matin2025teacher} and energy Hessians~\cite{amin2025towards}. 
Others aim to improve the accuracy of the training labels from teacher models utilizing ensemble-averaged forces from multiple teachers~\cite{ekstrom2023accelerating} or by first fine-tuning the teacher MLIPs with system-specific DFT data before distillation~\cite{gardner2025distillation,wang2025pre}.

Another shortcoming of most foundation MLIPs is the lack of explicit long-range electrostatics~\cite{kim2026reactive},
which may restrict the application to systems dominated by electrical response, such as ionic materials~\cite{Cheng2025Latent,zhong2025machine,li2026disentangling}, electrochemical interfaces~\cite{ King2025Machine,Wang2025Ionmodulateda,parker2026false,fan2026qnep}, liquid-vapor interfaces~\cite{niblett2021learning}, and redox processes in solution~\cite{kocer2024machine}. 
A number of methods~\cite{Baldwin2026,grasselli2026long,Kim2026} have been proposed to incorporate long-range interactions into MLIPs, including employing architectures or descriptors encoding global information~\cite{qu2026recipe, kosmala2023ewald, Caruso2026Extending, grisafi2019incorporating, huguenin2023physics},
or exploiting explicit atomic charges or Wannier centers~\cite{ko2021fourth,zhang2022deep,gao2022self}.

The Latent Ewald Summation (LES) method~\cite{Cheng2025Latent,King2025Machine,zhong2025machine,kim2025universal,Kim2026,kim2026polarizable} infers partial charges and resulting long-range electrostatics just from energy and force data. 
From the partial charges, LES further enables prediction of electrical response properties 
including Born effective charge (BEC) tensors and infrared (IR) spectra~\cite{King2025Machine, kim2025universal, Kim2026, Wang2025Ionmodulateda}.
However, LES has only been shown to capture electrostatics from DFT training sets,
and it is unclear whether the same is true in a distillation setting starting from a foundation MLIP.

Here we introduce a distillation workflow that trains lightweight LES-augmented MLIPs from foundation MLIPs.  
The premise is that long-range electrostatic response is implicitly encoded in the potential energy surface (PES): even though a foundation MLIP exposes only energies and forces and carries no explicit notion of charge, the electrostatic information present in its PES is sufficient for LES to learn. 
Concretely, for a specific system, configurations labeled by the energy and force information predicted by the foundation MLIP are used to train a student MLIP that combines a compact short-range MLIP baseline model such as MACE~\cite{batatia2022mace} or CACE~\cite{cheng2024cartesian} with the LES long-range augmentation.
The student therefore inherits the PES while gaining access to electrical-response properties, including Born effective charge tensors and infrared spectra, that the teacher cannot provide.

We considered a representative set of foundation models (Table~\ref{Table:foundation_models}), including UMA~\cite{wood2025umafamilyuniversalmodels}, MACE~\cite{Batatia2025foundationa,batatia2025crosslearning}, Orb~\cite{rhodes2025orbv3atomisticsimulationscale}, eSEN~\cite{sahoo2025open}, GemNet-OC~\cite{gasteiger2022gemnet,tran2023open}, PET~\cite{bigi2026pushing} and EquiformerV2-based~\cite{liao2023equiformerv2,allam2025aqcat25} models.
These foundation MLIPs differ mainly by the model architecture and the training dataset.
The main model choices include conservative potential energy surfaces versus direct force predictions, whether to enforce equivariance, cutoff ($r_{\mathrm{cut}}$) and the number of message-passing (MP) layers.
Training datasets for the tested foundation models span three main chemical domains: organic molecules (e.g., SPICE~\cite{eastman2023spice} V1/V2 with $\sim$1M $\omega$B97M-D3(BJ) configurations; OMol25~\cite{levine2025openmolecules2025omol25} with 102M $\omega$B97M-V configurations), 
inorganic bulk materials (e.g., MPtrj~\cite{deng2023chgnet} with 1.5M PBE/PBE+U configurations; OMat24~\cite{barros2026open} with 118M PBE+U configurations;
MatPES~\cite{kaplan2025foundational} with 0.4M PBE/r$^2$SCAN configurations), and heterogeneous catalyst surfaces (e.g., OC20~\cite{chanussot2021open} with 265M RPBE configurations; OC22~\cite{tran2023open} with 9.9M PBE configurations; OC25~\cite{sahoo2025open} with 7.8M RPBE-D3 configurations). 
In addition, we tested foundation MLIPs with cross-domain training~\cite{nascimento2026mixture}, including UMA models~\cite{wood2025umafamilyuniversalmodels} based on task-conditioned mixtures of linear experts, and the MACE-MH model~\cite{batatia2025crosslearning} based on a multi-head replay post-training strategy.

After the distillation, we evaluated the performance of student models for three systems: liquid water, concentrated hydrochloric acid, and anatase \ce{TiO2}(101)-water interface.
We address two central questions: The first is whether one can distill long-range electrostatic MLIPs from foundation models.
This can help address the short-range limitations of most foundation models while yielding lightweight student MLIPs with superior computational efficiency. 
The second is how the distillation capability varies across foundation model architectures and the underlying DFT levels of theory. 
This further serves as a direct probe of the physical fidelity encoded in the different foundation models, since the electrical response predictions can be benchmarked against DFT calculations or experimental measurements.
In addition, we further explore how to refine the distilled student MLIPs. For the \ce{TiO2}-water interface, we examine whether fine-tuning distilled student models on a modest amount of higher-level DFT data improves their physical accuracy regarding structural and spectroscopic predictions.

\section{Results}

\subsection{Workflow}
\begin{figure}
  \centering
\includegraphics[width=0.45\textwidth]{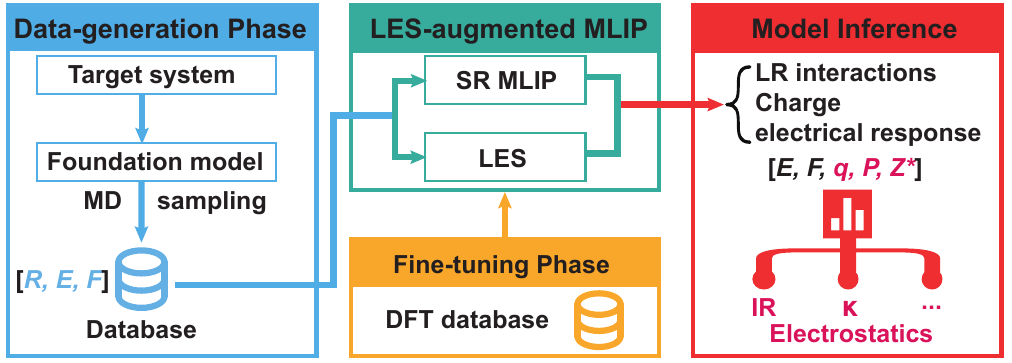}
  \caption{Schematic workflow for distilling long-range (LR) electrostatics from foundation models into a short-range (SR) MLIP integrated with the Latent Ewald Summation (LES) framework. The distilled model enables predictions of electrical response properties, including atomic partial charges $q$, polarizability $\mathbf{P}$, Born effective charge $\mathbf{Z}^*$, infrared (IR) spectra, and dielectric constant $\kappa$.  
  }
   \label{fig:workflow}
\end{figure}

\figref{fig:workflow} demonstrates our workflow for distilling knowledge from foundation models into lightweight MLIPs.
First, a training set with standard energy and force labels is generated from a chosen foundation model.
This can be achieved by conducting molecular dynamics (MD) simulations with the foundation model to sample the configurational space of the target system under certain thermodynamic conditions. 
Uncorrelated configurations ($\mathbf{R}$) are then extracted from the derived MD trajectories, together with their corresponding energies ($E$) and forces ($\mathbf{F}$), to form a training set.

Second, we fit a lightweight student MLIP augmented by LES to this synthetic training set to achieve computational efficiency and predict electrical properties including polarization ($\mathbf{P}$) and Born effective charges ($\mathbf{Z}^*$), which are not directly accessible from foundation MLIPs. 
More specifically, LES infers the latent charges $q_j^{\mathrm{les}}$ from local atomic environments with energy and force labels, and evaluates the $\alpha$ Cartesian component of $\mathbf{P}$ in the reciprocal $\mathbf{k}$-space for periodic systems~\cite{zhong2025machine}:
\begin{equation}
P_\alpha(\mathbf{k}) =
\sum_j \sqrt{\epsilon_{\infty}}
\frac{q_j^{\mathrm{les}}}{i \mathbf{k}}
\exp\left(i \mathbf{k}\cdot\mathbf{r}_j\right),
\label{eq:LES_P}
\end{equation}
where $j$ indexes atoms. Meanwhile, the Born effective charge tensor is then obtained in the long-wavelength limit:
\begin{equation}
Z_{j\alpha\beta}^*=
\lim_{\mathbf{k}\rightarrow \mathbf{0}} \operatorname{Re}\left[
\exp\left(-i\mathbf{k}\cdot\mathbf{r}_j\right)
\frac{\partial P_\alpha(\mathbf{k})}{\partial r_{j\beta}}
\right],
\label{eq:LES_Z}
\end{equation}
where $\alpha$ and $\beta$ denote Cartesian directions, $r_{j\beta}$ is the $\beta$ component of the position of atom $j$, and $\epsilon_\infty$ is the high-frequency dielectric constant accounting for electronic screening. 
Moreover, because foundation models may remain insufficiently accurate for system-specific quantum-chemical predictions~\cite{Batatia2025foundationa}, the distilled student models can be further refined using a small set of higher-level DFT energy and/or force labels for the target system.

\begin{table*}[htbp]
    \centering
    \caption{Summary of all foundation models evaluated, including architecture, conservative force constraint, maximum spherical degree for message-passing (MP) ($L_{\rm max}$), dataset, level of theory, cutoff, number of MP layers, and parameter count. ORB-v3 models, PET-OMATPES, and GemNet-OC22 are marked as N/A for $L_{\rm max}$ because spherical harmonics enter ORB-v3 only through edge featurization (vanilla), PET-OMATPES uses a rotationally unconstrained architecture, and GemNet-OC22 uses invariant scalar geometric representations.}
    \label{Table:foundation_models}
    \renewcommand{\arraystretch}{1.4}
    \setlength{\tabcolsep}{6pt}
    
    \resizebox{\textwidth}{!}{
\begin{tabular}{lllclccrc}
\hline
\textbf{Model}                                                                       & \textbf{Architecture}                                                                       & \textbf{Conservative} & $L_{\rm max}$   & \textbf{Dataset}                                                                                                                                                          & \textbf{Cutoff [\AA]} & \textbf{MP Layers} & \textbf{\# of Params} & \textbf{Charge/Spin} \\ \hline

UMA-S~\cite{wood2025umafamilyuniversalmodels}                  & \multirow{2}{*}{\begin{tabular}[c]{@{}l@{}}eSEN + Mixture of\\ Linear Experts\end{tabular}} & \multirow{2}{*}{Yes} & 2 & \multirow{2}{*}{\begin{tabular}[c]{@{}l@{}}OMol25 (main, $\omega$B97M-V), OMat24 (PBE), \\ OMC25 (PBE-D3),  ODAC25 (PBE-D3), OC20 (RPBE),\\ OC22 (PBE), OC25 (RPBE-D3) \end{tabular}}                       & \multirow{2}{*}{6.0}                     & 4                  & 146.6M                & Yes  \rule{0pt}{4ex}                \\
UMA-M~\cite{wood2025umafamilyuniversalmodels}                   &                                                                                             &                & 4            &                                                                                                                                                                           &                                          & 10                 & 1,396.7M              & Yes     \rule{0pt}{6.2ex}             \\ 
\hline
eSEN-OC25-sm~\cite{sahoo2025open}                              & eSEN (con.)                                                                               & Yes  & 2 & \multirow{2}{*}{OC25 (RPBE-D3)}                                                                                                                                           & \multirow{2}{*}{6.0}                     & 4                  & 6.3M                  & \multirow{2}{*}{No}  \\
eSEN-OC25-md~\cite{sahoo2025open}                              & eSEN (direct)       &   No           & 4              &                                                                                                                                                                           &                                          & 10                 & 50.7M                 &                      \\ \hline
MACE-OFF23(S)~\cite{kovacs2025MACEOFF}                         & \multirow{6}{*}{MACE}                                                                       & Yes              & 0   & SPICE V1 ($\omega$B97M-D3(BJ))                                                                                                                                               & 4.5                                      & 2                  & 0.7M                  & No                   \\
MACE-OFF23(M)~\cite{kovacs2025MACEOFF}                         &                                                                                             & Yes            & 1     & SPICE V1 ($\omega$B97M-D3(BJ))                                                                                                                                               & 5.0                                      & 2                  & 1.4M                  & No                   \\
MACE-OFF24(M)~\cite{kovacs2025MACEOFF}                         &                                                                                             & Yes           & 1     & SPICE V2 ($\omega$B97M-D3(BJ))                                                                                                                                               & 6.0                                      & 2                  & 1.4M                  & No                   \\
MACE-MP-0(L)~\cite{Batatia2025foundationa}                     &                                                                                             & Yes       & 2      & MPtrj (PBE/PBE+U)                                                                                                                                                             & 6.0                                      & 2                  & 5.7M                  & No                   \\
MACE-OMol~\cite{batatia2025crosslearning}                                                                            &                                                                                             & Yes        & 2         & OMol25 ($\omega$B97M-V)                                                                                                                                                     & 6.0                                      & 2                  & 52.4M                 & Yes                  \\
MACE-MH-1~\cite{batatia2025crosslearning}                      &                                                                                             & Yes      & 2           & \begin{tabular}[c]{@{}l@{}}OMat24 (main, PBE+U), OMol25 ($\omega$B97M-V), \\ OC20 (RPBE), SPICE ($\omega$B97M-D3(BJ)), \\ RGD1 (B3LYP), MPtrj (PBE+U), MatPES (r$^2$SCAN)\end{tabular} & 6.0                                      & 2                  & 13.8M                 & No                   \\ \hline
MACELES-OFF~\cite{kim2025universal}                            & MACELES                                                                              & Yes          & 1        & SPICE V1 ($\omega$B97M-D3(BJ))                                                                                                                                       & 4.5                                      & 2                 & 1.9M                & No                  \\ \hline 

Orb-v3-OMat~\cite{rhodes2025orbv3atomisticsimulationscale}     & ORB-v3 (con.)                                                                               & Yes       & N/A (Vanilla)         & OMat24 (PBE+U)                                                                                                                                                            & \multirow{3}{*}{6.0}                     & 5                  & 25.5M                 & No                   \\
Orb-v3-OrbMol~\cite{rhodes2025orbv3atomisticsimulationscale}   & ORB-v3 (con.)                                                                               & Yes       & N/A (Vanilla)         & OMol25 ($\omega$B97M-V)                                                                                                                                                     &                                          & 5                  & 25.8M                 & Yes                  \\
Orb-v3-OrbMol-d~\cite{rhodes2025orbv3atomisticsimulationscale} & ORB-v3 (direct)                                                                             & No      & N/A (Vanilla)              & OMol25 ($\omega$B97M-V)                                                                                                                                                     &                                          & 5                  & 25.9M                 & Yes                  \\ \hline 
PET-OMATPES~\cite{bigi2026pushing} & PET & Yes & N/A (Unconstrained) & OMat24 base (PBE) + MatPES fine-tuning (r$^2$SCAN) & 4.5 & 3 & 192.9M & No  \\ \hline

GemNet-OC22~\cite{tran2023open}                            & GemNet-OC                                                                                & Yes            & N/A (Invariant)    & OC22 (PBE)                                                                                                                                               & 6.0                                      & 4                 & 38.8M                & No                  \\ \hline 
AQCat25-ev2~\cite{allam2025aqcat25}                            & EquiformerV2                                                                                & No      & 6           & AQCat25 (RPBE), OC20 (RPBE)                                                                                                                                               & 6.0                                      & 20                 & 153.0M                & Yes                  \\ \hline
\end{tabular}
    }
\end{table*}

\subsection{Liquid Bulk Water}

\begin{figure*}
  \centering
\includegraphics[width=\textwidth]{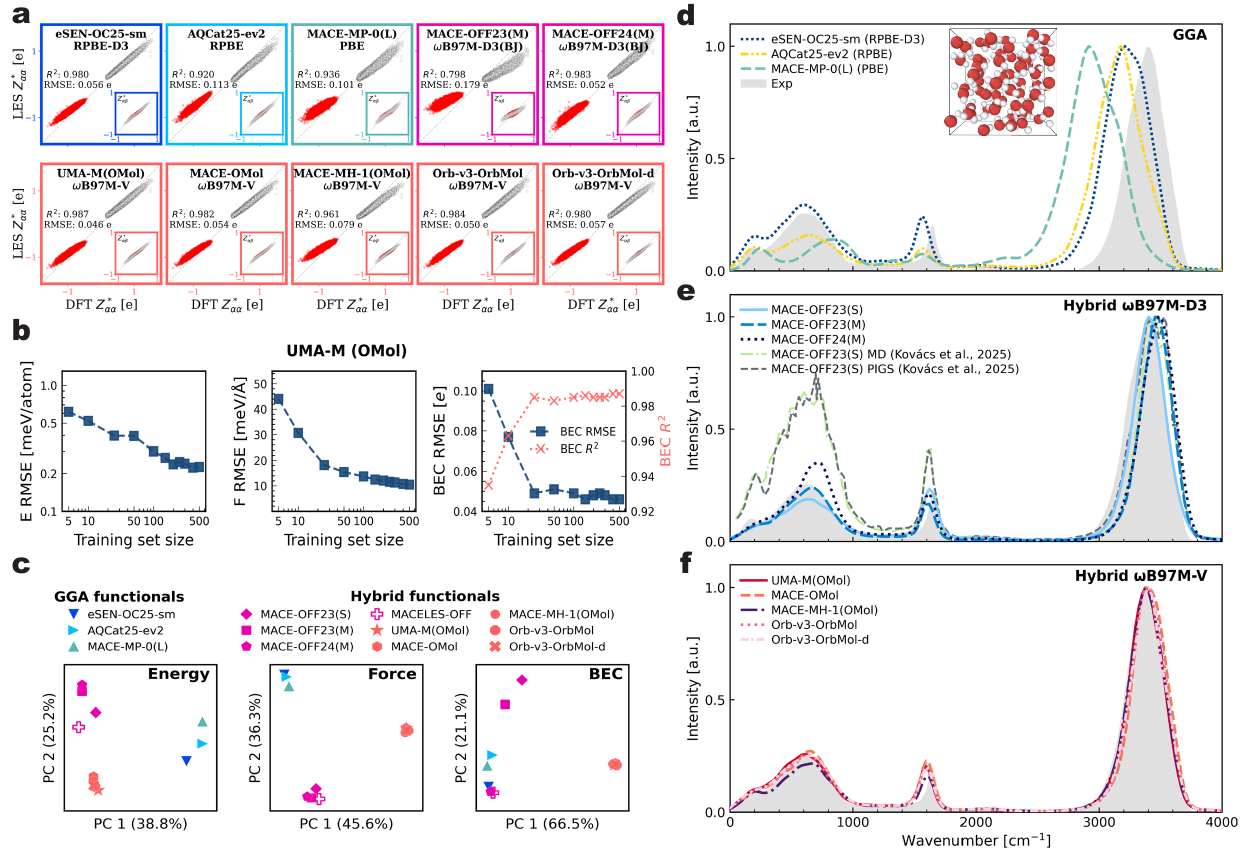}
  \caption{Benchmark results for distilled student MACELES models of liquid water derived from 11 foundation models.
  \textbf{a}: Parity plots of the predicted BEC tensor components of MACELES models against RPBE-D3 DFT reference values~\cite{Schmiedmayer2024} for 100 water configurations containing 64 molecules. 
  \textbf{b}: Learning curves for the MACELES models distilled from UMA-M with the OMol task for liquid water, which show the test root-mean-square errors (RMSEs) in energy, force, and BEC as a function of training-set size. 
   \textbf{c}: kPCA analyses of energies, forces, and BECs predicted by 10 student MLIPs on 100 test configurations. Each point represents one MLIP model, and the two axes are the kernel principal components. The directly trained long-range MACELES-OFF model~\cite{kim2025universal}, fitted to the SPICE V1 dataset at the $\omega$B97M-D3(BJ) level of theory, is included as a reference.
  \textbf{d-f}:  IR absorption spectra of bulk liquid water computed using student models grounded on different DFT levels of theory: \textbf{d}, GGA level; \textbf{e}, hybrid $\omega$B97M-D3(BJ); and \textbf{f}, hybrid $\omega$B97M-V. The experimental IR spectrum~\cite{Bertie1996Infrared} (shaded area) and previous PIGS results using MACE-OFF23(S)~\cite{kovacs2025MACEOFF} (dashed gray line) are included for comparison.
  }
   \label{fig:water}
\end{figure*}
  
An accurate description of bulk water is a critical benchmark for all foundation models, given its ubiquity in life and technology. 
Based on the workflow in \figref{fig:workflow},
we distilled a set of student MACELES models from 11 different foundation models, grouped into three categories based on the underlying DFT approximation:
three models based on generalized gradient approximation (GGA)-level functional (PBE~\cite{perdew1996generalized} for MACE-MP-0(L), RPBE~\cite{hammer1999improved} for AQCat25-ev2, and RPBE-D3~\cite{hammer1999improved,grimme2010consistent} for eSEN-OC25-sm), 
five models based on the hybrid-level functional $\omega$B97M-V/def2-TZVPD~\cite{mardirossian2016omegab97m,rappoport2010property,hellweg2015development} (UMA-M(OMol), MACE-OMol, MACE-MH-1(OMol), Orb-v3-OrbMol, Orb-v3-OrbMol-d),
and three models based on the hybrid-level functional $\omega$B97M-D3(BJ)/def2-TZVPPD~\cite{najibi2018nonlocal,weigend2005balanced,rappoport2010property,grimme2011effect,grimme2010consistent} (MACE-OFF23(S), MACE-OFF23(M), MACE-OFF24(M)).

We began by generating uncorrelated configurations of bulk liquid water via canonical (NVT) MD simulations of a 64-water cell using the UMA-M with the OMol task~\cite{wood2025umafamilyuniversalmodels}.  
The resulting configurations were labeled with energies and forces derived from UMA-M(OMol), yielding 500 training snapshots and 50 test snapshots for fitting MACELES student models.
The same 550 configurations were re-evaluated using another ten foundation models to construct the datasets for training student models. 
Moreover, to examine whether the structural ensemble was biased by the UMA-M(OMol) sampling procedure, we additionally distilled six student models using energies and forces recomputed by the corresponding foundation models on 654 configurations from previous RPBE-D3 \emph{ab initio} molecular dynamics (AIMD) simulations of liquid water~\cite{Schmiedmayer2024}.
As discussed in the Supplementary Information, student MLIPs trained on UMA MD trajectories yield BEC tensors comparable to those trained on AIMD trajectories. 
Details of the MD sampling, dataset construction, and MLIP training are provided in the Methods section.

\textbf{Born effective charges} 
To benchmark the electrostatics of the student MACELES models, we first assessed their BEC predictions, employing the experimental electronic dielectric constant ($\epsilon_\infty$=1.78)~\cite{leontyev2010electronic} in Eq.~\ref{eq:LES_P} and Eq.~\ref{eq:LES_Z} throughout.  
\figrefsub{fig:water}{a} compares the inferred BEC values with the RPBE-D3 DFT reference values~\cite{Schmiedmayer2024}.
\figrefsub{fig:water}{b} further demonstrates that BECs are learned with higher data efficiency compared to energy and force training. 
For the UMA-M(OMol)-distilled MLIPs, the BEC RMSE drops rapidly from $\sim$0.10~e to $\sim$0.05~e with only 25 configurations, while the $R^2$ values rise to $\sim$0.98 and remains nearly saturated thereafter.

The accuracy of distilled BEC tensors is sensitive to the receptive field of the teacher MLIP. 
\figrefsub{fig:water}{a} reveals that most student models accurately reproduce the DFT BECs with $R^2 > 0.9$, except for MACE-OFF23(M) with a shorter cutoff of 5.0~\AA{} and a receptive field of 10.0~\AA{}.
In comparison, the student model distilled from MACE-OFF24(M), which uses the same architecture but a larger 6.0~\AA{} cutoff, achieves substantially improved BEC accuracy ($R^2 = 0.98$). 
A similar trend is observed for two foundation models trained on the SPICE V1 dataset: BECs can be accurately distilled from the LES-augmented MACELES-OFF teacher model with a 4.5~\AA{} cutoff ($R^2=0.98$), whereas accuracy is much lower for the short-range MACE-OFF23(S) model with the same cutoff ($R^2=0.61$; see the Supplementary Information).
This comparison highlights the effectiveness of the LES framework in capturing long-range electrostatic interactions relative to conventional message-passing MLIPs.

To further assess the correlations among foundation models, we embedded their predictions for energies, forces, and BECs into a two-dimensional latent space using kernel principal component analysis (kPCA), as shown in \figrefsub{fig:water}{c}.
For each model $i$, the predicted energies, force components, or BEC tensor components were collected into separate flattened feature vectors $X_i$.
Similarity between model $i$ and $j$ is then measured by squared Euclidean distances ($d_{ij}^2 = \|X_i - X_j\|^2$). 
These distances are mapped into a matrix through a radial basis function kernel, $K_{ij} = \exp(-\gamma d_{ij}^2)$, where $\gamma$ is a tunable bandwidth hyperparameter. 
The resulting kPCA maps in \figrefsub{fig:water}{c} reveal that model predictions cluster primarily by their underlying DFT level of theory, rather than by architecture. 
Notably, MACE-OFF23(S) (pink diamond) and MACE-OFF23(M) (pink square) cluster closely with MACELES-OFF (pink plus) and MACE-OFF24(M) (pink pentagon) in the energy (left panel) and force (center panel) spaces, but not in the BEC space (right panel).
This divergence reflects the limited receptive field of MACE-OFF23(S) and MACE-OFF23(M) as discussed above, and thus demonstrates that BEC predictions can expose the long-range limitations in foundation MLIPs that are not apparent from energies and forces alone. 

\textbf{IR spectra}  
IR spectra provide a rigorous test of MLIP models since they are inherently sensitive to both intermolecular structure and intramolecular vibrations~\cite{hollas2004modern}. 
We computed the IR spectra for liquid water at 300~K at the experimental density of 0.997~g/cm$^{3}$~\cite{haynes2016crc}, and compared them with experiments~\cite{Bertie1996Infrared}. 

From the MD trajectory, the frequency-dependent IR spectrum, $I(\omega)$, is obtained via the Fourier transform of the current-current autocorrelation function~\cite{Schmiedmayer2024,zhong2025machine}: 
\begin{equation}
    I(\omega) \propto \frac{1}{T} \int_0^T dt \left\langle \mathbf{J}(0) \mathbf{J}(t) \right\rangle e^{-i\omega t},
    \label{eq:IR-spectra}
\end{equation}
where $\langle \cdots \rangle$ denotes an ensemble average over time origins and independent trajectories, and the time-dependent polarization current $\mathbf{J}(t)$ is defined as the sum over atomic velocities $\mathbf{v}_j(t)$ weighted by instantaneous BEC tensors $\mathbf{Z}_j^*(t)$, i.e., $\mathbf{J}(t) = \sum_{j=1}^N \mathbf{Z}_j^*(t) \mathbf{v}_j(t)$. 
In practice, we used the Wiener-Khinchin theorem~\cite{reichl2016modern} and a block-averaging scheme to compute the IR.
To facilitate a direct comparison of linewidths and spectral features, all IR intensities were normalized to the experimental peak maximum of the O–H stretching band~\cite{Bertie1996Infrared}, as shown in \figsrefsub{fig:water}{d-f}. 

To account for nuclear quantum effects (NQEs) on vibrational frequencies, all IR spectra computed from classical MD simulations in \figsrefsub{fig:water}{d-f} were corrected using empirical frequency-dependent redshifts: 5~$\mathrm{cm}^{-1}$ for the low-frequency translational and libration bands~\cite{Reddy2017, marsalek2017quantum}, 60~$\mathrm{cm}^{-1}$ for the bending band, and 175~$\mathrm{cm}^{-1}$ for the stretching band~\cite{moberg2018, savoj2024}. 
The validity of this correction is supported by the close agreement between the corrected classical IR spectrum from MACE-OFF23(S) (green dashed line in \figrefsub{fig:water}{d}) and the NQE-corrected IR spectrum (gray dashed line) from path integral coarse-grained simulations (PIGS) using the same model~\cite{kovacs2025MACEOFF}.

GGA-level models systematically redshift the intramolecular bending ($\sim$1650~cm$^{-1}$) and stretching ($\sim$3400~cm$^{-1}$) modes, as shown in \figrefsub{fig:water}{d}.
This redshift indicates that GGA-level functionals predict overly weakened O–H covalent bonds due to the larger delocalization errors inherent to these semi-local exchange-correlation functionals~\cite{pestana2017ab,marsalek2017quantum}, consistent with prior MLIP MD work~\cite{dao2026systematic}. 
Meanwhile, \figrefsub{fig:water}{d} and \figrefsub{fig:water}{f} show that IR spectra obtained from hybrid-level foundation models are in good agreement with the experimental reference (shaded region). 
This demonstrates the superior accuracy of the hybrid-level DFT functional with exact exchange relative to GGA-level descriptions.
Moreover, despite substantial architectural differences among the teacher foundation models (see Table~\ref{Table:foundation_models}), the corresponding student models yield consistent IR spectra. 
The remaining differences, such as those between MACE-OFF23(S) and MACE-OFF23(M) in \figrefsub{fig:water}{e}, can be attributed to the differences in receptive field which influence the accuracy of electrostatics distillation, as discussed above.
Overall, these IR spectra reveal that the underlying level of DFT theory plays a much more important role than model architecture in determining the electrical response properties of liquid water, consistent with the clustering observed in \figrefsub{fig:water}{c}.

\subsection{2~M HCl solution}

\begin{figure*}
  \centering
\includegraphics[width=0.7\textwidth]{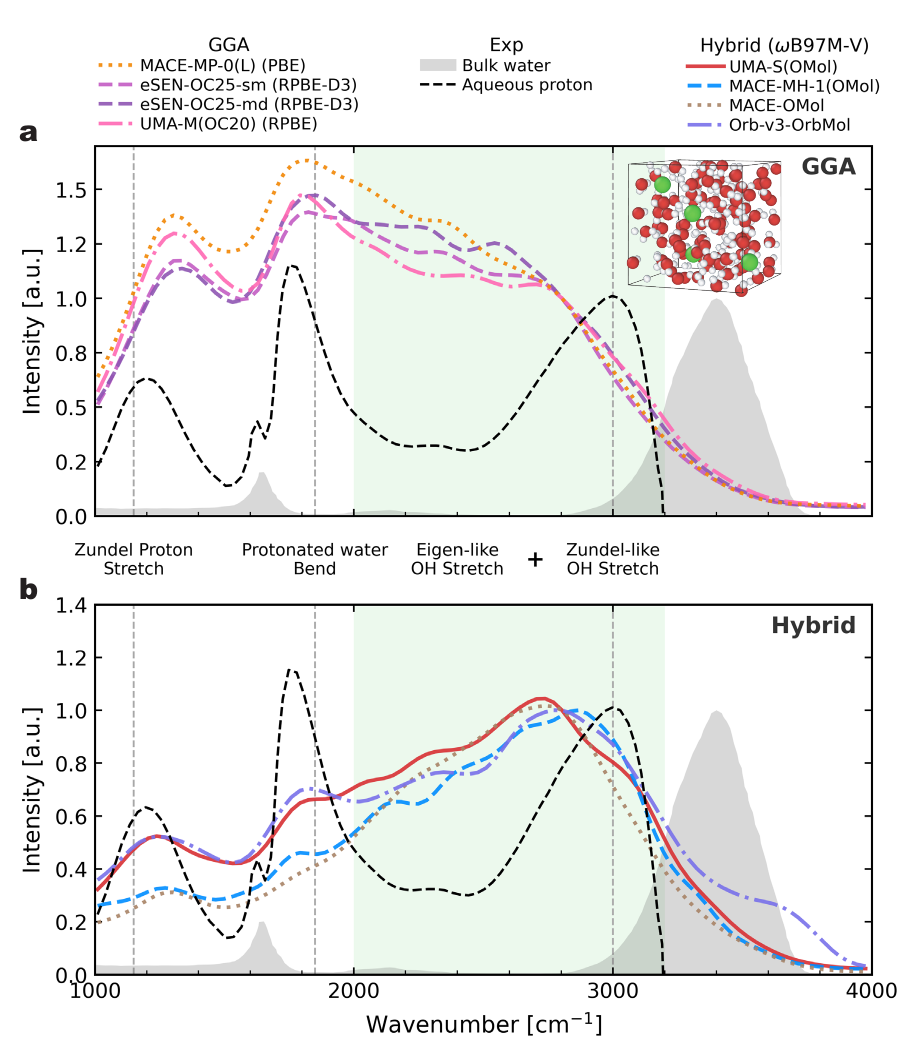}
  \caption{ Infrared (IR) difference spectra of the hydrated proton in 2~M HCl computed with student models based on: \textbf{a}) GGA-level functionals and \textbf{b}) hybrid-level functionals.
  The dashed black curve shows the aqueous proton spectrum measured from a 2~M HCl solution~\cite{fournier2018broadband}, while the gray shaded profile shows the experimental IR of pure water~\cite{Bertie1996Infrared}. 
  Vertical markers indicate characteristic vibrational signatures of the hydrated proton reported from previous experimental studies~\cite{fournier2018broadband}, specifically the Zundel proton stretch ($\sim$1150~cm$^{-1}$), protonated water bend ($\sim$1850~cm$^{-1}$), and the broad high-frequency O–H stretching continuum  involving both the Eigen-like and Zundel-like cations (2000-3200~cm$^{-1}$). 
  }
   \label{fig:2M_HCl}
\end{figure*}

We applied the distillation workflow to the concentrated hydrochloric acid (HCl) solution at 2~M concentration, given the significance of the aqueous excess proton in chemistry and biology~\cite{cukierman2006tu, agmon2016protons}.  
IR spectroscopy is a widely used probe for these systems due to its high sensitivity to coupled vibrational dynamics of the hydrated proton and the surrounding hydrogen-bond network~\cite{headrick2005spectral,thamer2015ultrafast,dahms2016hydrated},
but interpreting these spectra presents a major theoretical challenge~\cite{kulig2013clusters,brunig2022spectral,fournier2018broadband}. 

We collected 1,100 uncorrelated snapshots for 2~M HCl from NVT MD simulations using the UMA-S(OMol) V2 model.
The same configurations were then recomputed using eight foundation models and used to fit the corresponding MACELES student models. 
Four models considered here are based on GGA-level functionals (MACE-MP-0(L), eSEN-OC25-sm, eSEN-OC25-md, and UMA-M(OC20)),
and the other four models are based on the hybrid-level $\omega$B97M-V/def2-TZVPD method (UMA-S(OMol), MACE-MH-1(OMol), MACE-OMol, and Orb-v3-OrbMol). 
More details on the construction of the dataset and MLIP training are provided in the Methods section.

\textbf{IR spectra}  
We computed the IR difference spectra of the hydrated excess proton by separating the response of the protonated water complex from the background contribution of the surrounding neat water molecules. 
The total IR spectrum $I_{\rm tot}(\omega)$ for the entire 2~M HCl solution can then be computed as: 
\begin{equation}
\begin{aligned}
 I_{\rm tot}(\omega) \propto \left\langle \mathbf{J}_{\mathrm{H^+}}(0) \mathbf{J}_{\mathrm{H^+}}(t)\right\rangle
+ 2\left\langle \mathbf{J}_{\mathrm{H^+}}(0) \mathbf{J}_{\mathrm{water}}(t)\right\rangle  \\
+ \left\langle \mathbf{J}_{\mathrm{water}}(0)  \mathbf{J}_{\mathrm{water}}(t)\right\rangle ,
\end{aligned}
\label{Eq:HCl_IR_decomp}
\end{equation}
where $\mathbf{J}_{\mathrm{H^+}}(t)$ and $\mathbf{J}_{\mathrm{water}}(t)$ denote the polarization currents associated with the protonated water complex and the neat water molecules, respectively. 
The first two terms in Eq.~\ref{Eq:HCl_IR_decomp} define the IR difference spectrum of the hydrated proton, including the autocorrelation of the proton complex and its cross-correlation with the surrounding water. 
The contribution of neat water was computed from water molecules beyond the first solvation shell of the hydrated proton with a cutoff of 3.0~\AA{}, as discussed in the Supplementary Information.

Following previous experimental analyses~\cite{fournier2018broadband} as shown in \figref{fig:2M_HCl},  we characterized the IR difference spectrum of the hydrated excess proton in three regimes: 
1) the central Zundel proton-stretch mode near 1,200~cm$^{-1}$~\cite{dahms2017large, daly2017decomposition, biswas2017ir, headrick2005spectral};
2) the bending vibration of the protonated water complex around 1,750~cm$^{-1}$~\cite{thamer2015ultrafast, dahms2016hydrated, biswas2017ir, kim2002vibrational, xu2011infrared, vendrell2007full}; 
and 3) the broad O–H stretch absorption continuum of the hydrated proton spanning from 2,000 to 3,200~cm$^{-1}$ (green shaded region in \figref{fig:2M_HCl}), whose spectral profiles depend on whether the local solvation environment is more Eigen-like (\ce{H9O4^+}) or Zundel-like (\ce{H5O2^+})~\cite{thamer2015ultrafast, carpenter2018picosecond, biswas2017ir, xu2011infrared, fournier2018broadband}.

As shown in \figref{fig:2M_HCl}, the calculated IR difference spectra from classical MD simulations depend strongly on the underlying DFT functionals.
This sensitivity indicates that different DFT approximations yield markedly different coupling strength between the excess-proton defect and the surrounding hydrogen bond network. 
In particular, the GGA-level models in \figrefsub{fig:2M_HCl}{a} overestimate the intensities of both the Zundel stretching band near 1200~cm$^{-1}$ and the protonated-water bending band near 1750-1850~cm$^{-1}$. 
Consistent with the behavior observed for pure water (\figsrefsub{fig:water}{d}), these GGA-level MLIPs also produce a broader and more red-shifted O–H stretching response. 
These features can be explained by the weakened O–H covalent bonds and thus the more rigid hydrogen-bond network predicted by the GGA-level functionals. 
As further discussed in the Supplementary Information, the radial distribution functions obtained from the GGA models also exhibit a more structured second-shell region and indicate stronger intermediate-range ordering of the hydrogen-bond network.

In contrast, the hybrid-level student models yield reduced intensities for the proton-related stretching and bending vibrations, together with more localized O–H stretch peaks, as shown in \figrefsub{fig:2M_HCl}{b}.
Moreover, compared to the GGA-level results in \figrefsub{fig:2M_HCl}{a}, these spectral features are in better agreement with the experimental measurements~\cite{fournier2018broadband}, particularly in the high-frequency O–H stretching region. 
The improved agreement reflects the role of exact exchange in reducing the over-delocalization in semi-local functionals, and thus leads to a softer solvation shell around the hydrated proton. 

Overall, the predicted IR spectra are governed primarily by the underlying DFT level of theory.
Even when the teacher foundation MLIPs differ substantially in architecture and in the configurational space represented by their training sets, such as MACE-MP-0(L) and eSEN-OC25-sm, student models distilled from GGA-level labels yield similar IR difference spectra of the hydrated proton, as shown in \figrefsub{fig:2M_HCl}{a}. 
Similarly, student models distilled from foundation MLIPs trained on the same OMol25 datasets produce closely aligned IR difference spectra (\figrefsub{fig:2M_HCl}{b}).

\subsection{\ce{TiO2}-water interface}

\begin{figure*}
  \centering
\includegraphics[width=\textwidth]{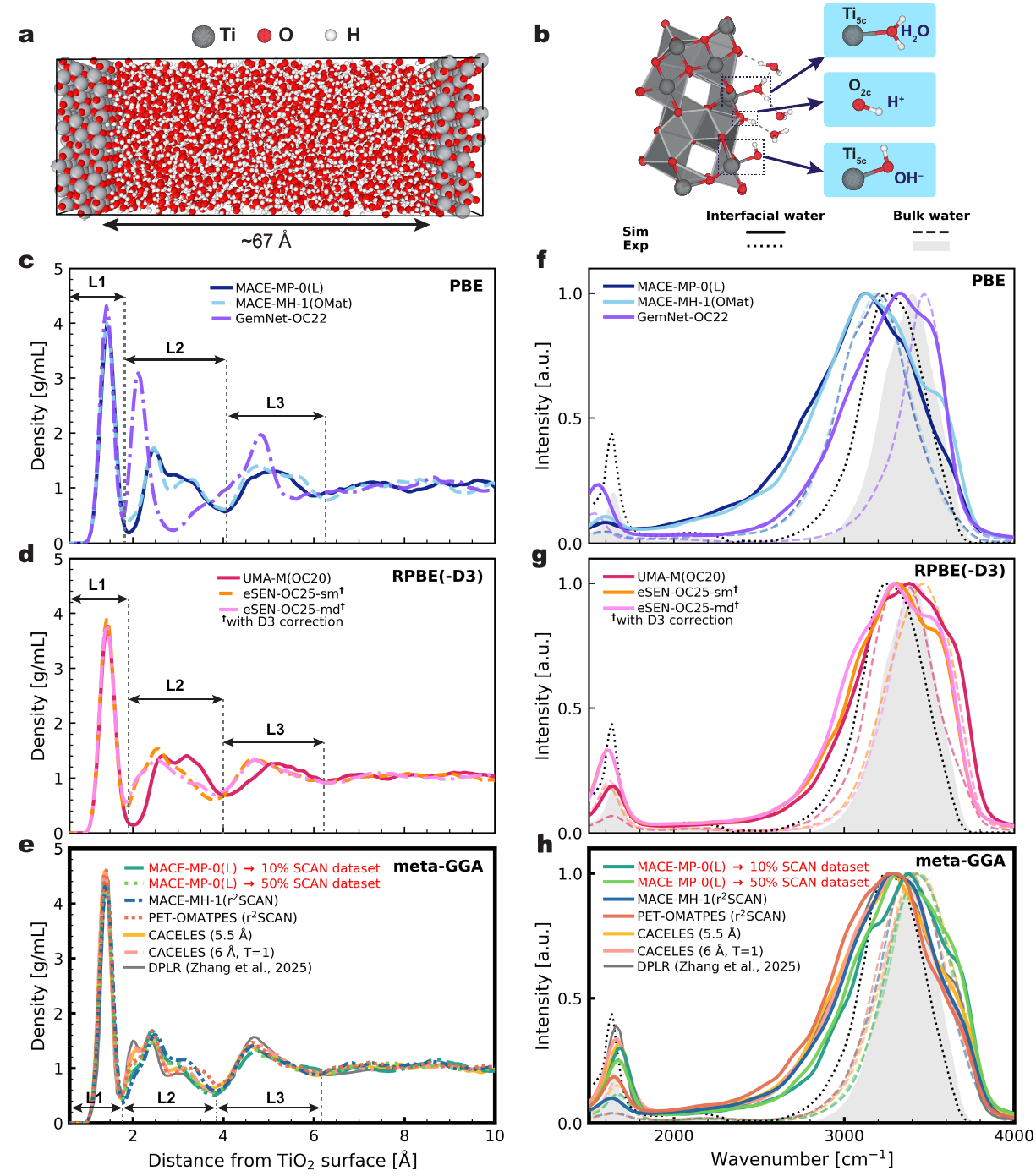}
  \caption{Structures and IR spectra of water at the anatase \ce{TiO2}(101)-water interface.
    \textbf{a}: Representative MD snapshot of the simulation cell ($30.7 \times 33.9 \times 83.4$~\AA$^3$) with 2376 \ce{H2O} molecules and 540 \ce{TiO2} formula units.
    \textbf{b}: The adsorbed species in the first layer include water molecules, protons, and hydroxide ions.
    \textbf{c-e}, Water density profiles as a function of distance from the \ce{TiO2} surface, grouped by the underlying DFT functional, with three interfacial layers marked as L1, L2, and L3.
        \textbf{c}, PBE-level MLIPs, including MACE-MP-0(L), MACE-MH-1(OMat), and GemNet-OC22;
        \textbf{d}, RPBE-level MLIPs, including UMA-M(OC20), eSEN-OC25-sm$^{\dagger}$, and eSEN-OC25-md$^{\dagger}$, where $\dagger$ indicates the inclusion of D3 dispersion corrections;
        \textbf{e}, meta-GGA-level models, including student models distilled from MACE-MH-1(r$^2$SCAN) and PET-OMATPES, MACE-MP-0(L) student models fine-tuned on 10\% and 50\% of the SCAN DFT dataset, directly fitted CACELES models with different hyperparameters ($r_{\mathrm{cut}}=5.5$~\AA{} with no message passing, and $r_{\mathrm{cut}}=6$~\AA{} with one message passing layer $T=1$), and the previously reported DPLR result~\cite{zhang2025_IR}.
    \textbf{f-h}, IR spectra of first-layer interfacial water, grouped by the same DFT functional categories as in \textbf{c-e}: 
        \textbf{f}, PBE; 
        \textbf{g}, RPBE (UMA-M(OC20)) and RPBE-D3 (eSEN-OC25 models); 
        and \textbf{h}, meta-GGA.
    For each model, colored solid and dashed curves show the simulated spectra of interfacial water and water more than 7~\AA{} from the interface, respectively.
    The black dotted curve is the experimental IR spectrum of interfacial water~\cite{connor1999infrared}, and the gray shaded profile is the experimental IR spectrum of bulk water~\cite{Bertie1996Infrared}.    
    }
   \label{fig:TiO2_water}
\end{figure*}

Metal oxide-electrolyte interfaces are critical to energy conversion, photocatalysis, and environmental chemistry~\cite{rousseau2020theoretical}. 
More specifically, the anatase \ce{TiO2}(101)-water interface stands out for its utility in electrocatalytic water splitting, photovoltaics, and anticorrosion coatings~\cite{schneider2014understanding, guo2019single,selloni2024aqueous}. 
To circumvent the high computational cost of AIMD simulations, recent theoretical studies have increasingly employed MLIPs to simulate the \ce{TiO2}-water interface, including short-range MLIPs for water adsorption and dissociation processes~\cite{wen2023water,zeng2023mechanistic,ling2026effect} and long-range deep potential (DPLR) models for electric double layers~\cite{zhang2024molecular} and IR spectrum of interfacial water~\cite{zhang2025_IR}.
Here, we applied the distillation workflow to evaluate whether foundation models can provide a computationally efficient and physically reliable route for exploring the structure and IR spectrum of interfacial water on \ce{TiO2}. 

The training set contains 3,346 \ce{TiO2}(101)-water interface configurations from previous AIMD simulations performed at the meta-GGA SCAN level~\cite{zhang2024molecular}. 
First, we trained CACELES models directly on this SCAN dataset, to validate the cutoff and message-passing settings, as detailed in the Methods section.
Then, we recomputed energies and forces for the training configurations using three groups of foundation models based on their underlying DFT level: 
PBE-based models including MACE-MP-0(L), MACE-MH-1(OMat), and GemNet-OC22; 
RPBE-based models including UMA-M(OC20), eSEN-OC25-sm, and eSEN-OC25-md; 
and r$^2$SCAN-based models including MACE-MH-1(r$^2$SCAN) and PET-OMATPES.
We fitted all the relabeled datasets to the student CACELES models.

\textbf{Water density and IR from distilled MLIPs} 
To benchmark the distilled student models, we computed water density profiles perpendicular to the \ce{TiO2}-water interface, from MD simulations at 330~K (see \figrefsub{fig:TiO2_water}{a}).
As shown in \figsrefsub{fig:TiO2_water}{c, d, e},
all student models predict a sharp first peak at $\sim$1.4~\AA{} corresponding to water molecules directly adsorbed at the undercoordinated \ce{Ti_{5c}} and \ce{O_{2c}} sites, followed by the second and third peaks that gradually decay to the bulk density beyond $\sim$6.5~\AA{}.

The computed interfacial water structure strongly depends on the DFT underlying the foundation model.
Although the r$^2$SCAN-level teacher models considered (MACE-MH-1(r$^2$SCAN) and PET-OMATPES) are trained exclusively on the crystalline material MatPES dataset without interfaces, the predicted water density profiles are qualitatively consistent with the previous SCAN-level DPLR simulations~\cite{zhang2025_IR}, demonstrating their out-of-the-box transferability.
Likewise, the PBE-level student models predict similar adsorption profiles (\figrefsub{fig:TiO2_water}{c}), while the teacher MACE-MP-0(L) was trained on MPtraj and MACE-MH-1(OMat) was trained on OMat24.  
However, the student model from GemNet-OC22 (purple dash-dotted line) stands as an outlier in the positions and intensities of the second and third water layer. 
Such deviation does not seem to come from the distillation procedure, 
as the identical protocol was followed and the training force errors are not abnormally large.
As such, the deviation may come from the GemNet-OC22 architecture or the training set.

\figsrefsub{fig:TiO2_water}{f, g, h} show the IR spectra without NQE corrections for first-layer water,
where the first layer is defined by the first minimum following the first peak in the water density profile and includes molecular water and dissociated species (\figrefsub{fig:TiO2_water}{b}). 
We also computed IR spectra for water molecules located more than 7~\AA{} from the interface (dashed lines in \figsrefsub{fig:TiO2_water}{f, g, h}), where the water density has relaxed to the bulk value. 
Experimental IR data for bulk water (gray-shaded region)~\cite{Bertie1996Infrared} and interfacial water (black dotted line)~\cite{connor1999infrared} are included for reference.

Across all distilled models, interfacial water (solid lines in \figsrefsub{fig:TiO2_water}{f, g, h}) exhibits a broader and red-shifted O–H stretching band relative to bulk water (dashed lines). 
This spectral broadening reflects the heterogeneous hydrogen-bonding environments at the oxide surface, where molecular water, hydroxyl groups, and protonated water complexes contribute to a wide distribution of O–H bond strengths, consistent with previous experimental measurements~\cite{o2023elucidating} and DPLR MD simulations~\cite{zhang2025_IR}. 

The surface IR spectra predicted by the student models are governed primarily by the DFT functional underlying their teachers.
The PBE-level IR spectra (\figrefsub{fig:TiO2_water}{f}) show larger redshifts for both bending and stretching modes, consistent with overstructured hydrogen bonding. 
The RPBE-level models (\figrefsub{fig:TiO2_water}{g}) capture qualitative features of interfacial water more accurately, but still deviate from the experimental IR measurements (black dotted line). 
With the improved meta-GGA level of theory (\figrefsub{fig:TiO2_water}{h}), both the r$^2$SCAN-level student models and the SCAN-level directly fitted models exhibit smaller O–H bending and stretching redshifts, in better agreement with experiment~\cite{connor1999infrared} and consistent with previous AIMD and MLIP MD studies showing that the meta-GGA functional better describes the \ce{TiO2}-water interface~\cite{nadeem2018water,calegari2018structure,zhang2025_IR}.

\textbf{Water density and IR from fine-tuned MLIPs}
To improve the accuracy of GGA-level student models, we fine-tuned the student model distilled from MACE-MP-0(L) using different fractions of the SCAN force data. 
Fine-tuning with 10\% of the dataset ($\sim$300 snapshots) nearly matches the force accuracy of directly trained SCAN MLIPs (see the Supplementary Information).

The fine-tuned models yield improved structural (\figrefsub{fig:TiO2_water}{e}) and spectroscopic predictions (\figrefsub{fig:TiO2_water}{h}), compared to the baseline student model. 
In particular, fine-tuning with only 10\% of the SCAN dataset reproduces a water density profile (solid green line) close to those from directly fitted SCAN CACELES models (orange and pink lines) and previous DPLR results (gray line)~\cite{zhang2025_IR}. 
Increasing the fine-tuning fraction to 50\% (dotted green line) yields only marginal additional improvement. 
For the IR spectra, both the SCAN-level CACELES and fine-tuned models reproduce the two main spectral signatures of surface IR spectra identified in the previous DPLR study~\cite{zhang2025_IR}.
First, surface binding and orientational ordering of first-layer water enhance the bending peak intensity near $\sim$1650~cm$^{-1}$ relative to bulk water (dashed lines), which is consistent with previous SCAN AIMD results~\cite{calegari2018structure}. 
Second, heterogeneous hydrogen bonding to \ce{O_{2c}} sites and surface hydroxyls weakens the O–H bonds of surface water~\cite{selloni2024aqueous}, resulting in a broadened, red-shifted O–H stretching band as discussed above for the GGA-level models. 
Overall, these spectra derived from the meta-GGA-level models agree better with the experimental interfacial-water spectrum (black dotted line) than those from the GGA-level distilled models.

Compared to previous DPLR models utilizing Wannier centers~\cite{zhang2024molecular,zhang2025_IR}, our LES-based distillation and fine-tuning strategy substantially reduces the cost of DFT calculations for dataset construction. 
At the same time, our workflow also eliminates the need for Wannier-center labels to learn long-range electrostatics while retaining quantitative structural (\figrefsub{fig:TiO2_water}{e}) and infrared (\figrefsub{fig:TiO2_water}{h}) predictions.
Similar to the previous DPLR work~\cite{zhang2024molecular}, our distilled or fine-tuned electrostatic MLIPs can also be used to simulate the differential capacitance, surface charging, and water dissociation reactions of this electrochemical interface under external electric fields.

\section{Discussion}
We show that LES-augmented MLIPs can learn long-range electrostatics, 
by training solely on energies and forces derived from foundation models.
The electrostatic distillation is data-efficient, as evidenced by the rapid saturation of the learning curve for BEC tensors within tens of configurations (\figrefsub{fig:water}{b}).
This work therefore extends the LES framework beyond its original scope that relies on DFT energies and forces~\cite{Cheng2025Latent,King2025Machine,kim2025universal,Wang2025Ionmodulateda} to simulate electrical properties for diverse chemical systems starting from foundation MLIPs.

The DFT level of theory employed in the teacher foundation models is the primary determinant of the distilled electrical response.
This is evidenced by kPCA analyses of energies, forces, and BEC tensors for bulk water (\figrefsub{fig:water}{c}) that illustrate student models cluster by DFT functional family rather than by architecture. 
Moreover, student models within the same DFT functional family generally yield consistent IR spectra for liquid bulk water, protonated water, and interfacial water (\figsrefsub{fig:water}{d-f}, \figref{fig:2M_HCl}, and \figsrefsub{fig:TiO2_water}{f-h}). 
In particular, hybrid-level models reduce the over-delocalization issue of GGAs and bring the O–H stretching spectra into better agreement with experiment for both bulk water (\figsrefsub{fig:water}{e, f}) and protonated water (\figrefsub{fig:2M_HCl}{b}). 

The architecture of the teacher foundation MLIP seems to play a minor role in distillation, except when the receptive field is too small.
When trained on the same dataset, foundation model architecture has only a small effect on spectroscopic observables, as shown by the consistent water IR spectra (\figrefsub{fig:water}{f}) and hydrated-proton IR difference spectra (\figrefsub{fig:2M_HCl}{b}) obtained from different OMol25-trained models.
As evidenced by the consistent IR spectra obtained from Orb-v3-OrbMol and Orb-v3-OrbMol-d for water, and from eSEN-OC25-sm and eSEN-OC25-md for 2~M HCl and \ce{TiO2}-water,  energy conservation and equivariance, or their absence, in the teacher foundation MLIP have no notable effect on the learned electrostatics.
In contrast, the receptive field of the teacher models can become the limiting factor in picking up sufficient information on electrostatic interactions.
As an example, the two outliers in the BEC clustering (\figrefsub{fig:water}{c}), MACE-OFF23(S) and MACE-OFF23(M), both have insufficient receptive fields, whereas the MACE-OFF24(M) model with a larger cutoff (6~\AA{}) yields substantially improved BEC predictions.

The distillation workflow also probes and validates the out-of-the-box transferability of foundation models to chemical environments not directly covered in their training set.
For the \ce{TiO2}–water interface, student MLIPs based on foundation models trained predominantly on crystalline inorganic materials, such as MACE-MH-1(r$^2$SCAN) and PET-OMATPES, reproduce a three-layered water density profile and interfacial IR spectra consistent with meta-GGA-level reference calculations (\figsrefsub{fig:TiO2_water}{e, h}). 
On the other hand, distillation can also expose limitations of foundation models: the GemNet-OC22-distilled student predicts anomalously overstructured second- and third-layer density peaks (\figrefsub{fig:TiO2_water}{c}). This calls for caution when applying foundation models to a target system.

Finally, fine-tuning student models with higher-level DFT data is a data-efficient route to improve predictive accuracy.
For the \ce{TiO2}-water interface, fine-tuning the PBE-level student model with only $\sim$300 SCAN configurations was sufficient to recover SCAN-level water density profiles (\figrefsub{fig:TiO2_water}{e}) and interfacial IR spectra (\figrefsub{fig:TiO2_water}{h}) that are consistent with previous DPLR results~\cite{zhang2025_IR}.
This data efficiency indicates that foundation model distillation can provide a low-cost starting point for simulations of complex systems, with the subsequent fine-tuning correcting residual system-specific errors.

A limitation of this work is that, for each tested system, student models were trained and evaluated on the same configurations generated by one selected foundation model. 
This protocol enables a controlled comparison of energy and force labels across foundation MLIPs, but may introduce a bias.
However, for bulk water we tested the influence of using a different set of training configurations collected from RPBE-D3 AIMD~\cite{Schmiedmayer2024},
and found that student models trained on configurations sampled from AIMD and UMA-M(OMol) MD yield consistent BEC tensors and IR spectra for liquid water, as discussed in the Supplementary Information.
Another limitation is that NQEs were considered only through a uniform empirical correction to the water IR spectra and were not included for protonated or interfacial water. 
Because NQEs can affect water vibrational dynamics differently across chemical environments and DFT functionals, future work should examine their role across bulk, protonated, and interfacial water for each DFT functional.

To conclude, we show that distilling foundation MLIPs into lightweight, long-range models is a viable strategy for simulating electrical response properties for bulk and interfacial systems,
with fine-tuning as an option to further improve predictions.
The distillation process largely preserves the accuracy and limitations encoded in the underlying DFT levels of foundation MLIPs, while also serving as a diagnostic of physical long-range interactions captured in these models.
Our work thus partially resolves the short-range limitations of most foundation models,
and broadens the scope of these models for predicting electrical response properties of diverse systems.

\section{Methods}

\subsection{Details on dataset construction}

\textbf{Liquid water}
NVT MD simulations of bulk liquid water were performed using UMA-M(OMol) in FAIRChem (V2)~\cite{wood2025umafamilyuniversalmodels}.
The simulation cell contained 64 water molecules at the experimental density of 0.997~g/cm$^{3}$~\cite{haynes2016crc}, with the temperature controlled at 300~K using a Nos\'e-Hoover thermostat.
Following a 40~ps pre-equilibration stage with a 1~fs timestep, 550 uncorrelated snapshots were extracted from a 1~ns production run, with 500 used for training and 50 for testing.
To verify that configurational sampling was not biased by the choice of foundation MLIPs, we trained six student models on an independent RPBE-D3 dataset of liquid water~\cite{Schmiedmayer2024} that consists of 604 training and 50 test configurations generated from multiple canonical RPBE-D3 AIMD heating and cooling trajectories of 64 water molecules at the same experimental density, spanning 270-420~K.

\textbf{2~M HCl solution}
We performed NVT MD simulations of 2~M HCl solution with the UMA-S(OMol) V2 model using a simulation cell containing 96 \ce{H2O} molecules and 4 \ce{HCl} pairs at 1.03~g/cm$^{3}$~\cite{haynes2016crc}, corresponding to the experimental density at room temperature. 
To sample a broader configurational space, the simulation temperature was elevated to 400~K.
Following a 100~ps pre-equilibration stage with a 1~fs timestep, we extracted 1,100 uncorrelated snapshots from a 2~ns trajectory to construct the dataset.

\textbf{\ce{TiO2}-water interface}
The training dataset comprised 3,346 anatase(101)-water interface structures extracted from a previous SCAN DFT dataset~\cite{zhang2024molecular} collected using active learning.
The system sizes of these structures range from 219 to 426 atoms, with the largest interface cells containing 60 \ce{TiO2} formula units and 82 \ce{H2O} molecules.  

\subsection{Details on MLIP training}

All student models use the LES augmentation with a one-dimensional latent charge variable, $\sigma = 1$~\AA{} and $dl=2$~\AA{} ($k_c = \pi$). The training and test errors of all fitted models are summarized in the Supplementary Information.

\textbf{Liquid water} 
The student models were trained with the LES-augmented MACE architecture with $r_{\mathrm{cut}} = 4.5$~\AA{}, $L_{\rm max}=1$, and two MP layers (\texttt{num\_interactions}=2). 
The MACELES model contains $\sim$0.18~M parameters, about three to four orders of magnitude fewer than the teacher models.
All student models yield comparable training and test root-mean-square errors (RMSEs), with most achieving test energy errors below 0.5~meV/atom and test force errors below $\sim$25~meV/\AA{}.

\textbf{2~M HCl solution}
The student models for the 2~M HCl solution used the MACELES architecture with $r_{\mathrm{cut}} = 5.5$~\AA{}, $L_{\rm max}=1$, and two MP layers.  
Each distilled MACELES model contains $\sim$0.19~M parameters.
All student models exhibit low and comparable training and test errors, with test RMSEs remaining below 0.7~meV/atom for energies and below 20~meV/\AA{} for forces, and yield consistent BEC predictions for bulk water, as discussed in the Supplementary Information.

\textbf{\ce{TiO2}-water interface} 
The student models for the \ce{TiO2}-water interface employed a LES-augmented CACE (CACELES) architecture with $r_{\mathrm{cut}} = 5.5$~\AA{}, 6 Bessel radial functions, $c=12$, $\ell_{\max }=3$, $\nu_{\max }=3$, $N_{\mathrm{embedding}}=5$, and no MP layer ($T=0$). 
The training and test errors are comparable across all student and directly fitted models, with test RMSEs generally below 0.3~meV/atom for energies and below 80~meV/\AA{} for forces.

To validate the cutoff and message-passing settings, we trained a CACELES model directly on the SCAN dataset and compared its performance with another directly fitted model using a larger cutoff ($r_{\mathrm{cut}} = 6.0$~\AA{}) and one MP layer ($T=1$).
Although the CACELES model with $T=1$ gives a slightly lower test force RMSE than the one with $T=0$ (67.1 versus 73.3~meV/\AA{}) and contains more parameters ($\sim$0.10~M versus $\sim$0.05~M), the two models predict consistent water density profiles and surface IR spectra as shown in \figsrefsub{fig:TiO2_water}{e, h}.

\subsection{Details on MLIP MD simulations}

\textbf{Liquid water} 
NVT MD simulations were performed at 300~K using a Nos\'e-Hoover thermostat in a simulation cell with 64 water molecules at the experimental density of 0.997~g/cm$^{3}$~\cite{haynes2016crc}.
For each student model, a 10~ps pre-equilibration stage was followed by a 50~ps production MD run in the Atomic Simulation Environment (ASE)~\cite{hjorth_larsen_atomic_2017}.
A timestep of 0.25~fs was used to better sample high-frequency vibrational dynamics. 

\textbf{2~M HCl solution}
NVT MD simulations were performed at 300~K using a Nos\'e-Hoover thermostat in a cell containing 96 \ce{H2O} molecules and 4 \ce{HCl} pairs at the experimental density of 1.03~g/cm$^{3}$~\cite{haynes2016crc}.
After 50~ps of pre-equilibration, a 50~ps production trajectory was generated in ASE with a timestep of 0.25~fs.

\textbf{\ce{TiO2}-water interface} 
To minimize finite-size effects and maintain consistency with previous work~\cite{zhang2025_IR}, here we conducted NVT MD simulations with a Nos\'e-Hoover thermostat for \ce{TiO2}(101)-water interfaces using an orthorhombic simulation cell of 8,748 atoms ($30.7 \times 33.9 \times 83.4$~\AA$^3$) (\figrefsub{fig:TiO2_water}{a}). 
In particular, the \ce{TiO2} substrate is a five-layer $(3 \times 9)$ anatase(101) slab comprising 540 \ce{TiO2} formula units, with the Cartesian $x$, $y$, and $z$ axes aligned along the $[10\bar{1}]$, $[010]$, and $[101]$ crystallographic directions, respectively.
The water region extends approximately 67~\AA{} along the surface normal and separates the periodically repeated slabs.
Each of the two equivalent \ce{TiO2} surfaces exposes 54 twofold-coordinated oxygen sites, \ce{O_{2c}}, and 54 fivefold-coordinated titanium sites, \ce{Ti_{5c}}.

For each student model, we first performed a 1.0~ns pre-equilibration simulation at 450~K to accelerate slow interfacial rearrangements and proton-transfer processes.
During this stage, the hydrogen mass was increased to 10~amu, and a timestep of 1.0~fs was used.
Production simulations were then performed at 330~K with a timestep of 0.3~fs for 60~ps, consistent with the setups used in previous MLIP MD simulations of the same interface~\cite{zhang2025_IR}.

\textbf{Acknowledgments}
Research reported in this publication was supported by the National Institute of General Medical Sciences of the National Institutes of Health under Award Number R35GM159986.
The content is solely the responsibility of the authors and does not necessarily represent the official views of the National Institutes of Health.
The authors acknowledge the research computing facilities provided by BRC UCB and LRC Lawrence Berkeley National Laboratory. 

\textbf{Data availability statement}
Training data, fitted MLIPs, and data analysis scripts generated for the study are in the Supplementary Information repository: \url{https://github.com/ChengUCB/les_distill}.

\textbf{Code availability} 
The LES library is publicly available at 
\url{https://github.com/ChengUCB/les}.
The MACE package with the LES implementation is available at \url{https://github.com/acesuit/mace}.
The CACE package with the LES implementation is available at \url{https://github.com/BingqingCheng/cace}.

\bibliography{refs}

\end{document}